\documentclass{emulateapj}

\shorttitle{Radio AGN population dichotomy}
\shortauthors{V. Smol\v{c}i\'{c}}

\def\f#1   {Fig.~\ref{#1}}
\def\s#1   {Sec.~\ref{#1}}
\def\tab#1   {Tab.~\ref{#1}}
\def\t#1   {Tab.~\ref{#1}}
\def\lum   {$\mathrm{L}_\mathrm{1.4GHz}$}
\def\comm#1   {{\tt (COMMENT: #1) }}

\def\msol              {$\mathrm{M}_{\odot}$}

\def\wh                {W~Hz$^{-1}$}

\slugcomment{  }

\begin{document}

\title{ The radio AGN population dichotomy: \\ Green valley Seyferts versus red
  sequence low-excitation AGN}

\author{V.~Smol\v{c}i\'{c}\altaffilmark{1}
        }
\altaffiltext{1}{ California Institute of Technology, MC 105-24, 1200 East
California Boulevard, Pasadena, CA 91125 }

\begin{abstract}
Radio outflows of active galactic nuclei (AGN) are invoked in
cosmological models as a key feedback mechanism in the latest phases
of massive galaxy formation. Recently it has been suggested that the
two major radio AGN populations -- the powerful high-excitation, and
the weak low-excitation radio AGN (HERAGN and LERAGN, resp.) --
represent two earlier and later stages of massive galaxy build-up. To
test this, here we make use of a local ($0.04<z<0.1$) sample of
$\sim500$ radio AGN with available optical spectroscopy, drawn from
the FIRST, NVSS, SDSS, and 3CRR surveys.  A clear dichotomy is found
between the properties of low-excitation (absorption line AGN, and
LINERs) and high-excitation (Seyferts) radio AGN. The hosts of the
first have the highest stellar masses, reddest optical colors, and
highest mass black holes but accrete inefficiently (at low rates). On
the other hand, the high-excitation radio AGN have lower stellar
masses, bluer optical colors (consistent with the `green valley'), and
lower mass black holes that accrete efficiently (at high rates). Such
properties can be explained if these two radio AGN populations
represent different stages in the formation of massive galaxies, and
thus are also linked to different phases of the `AGN feedback'.

\end{abstract}


\section{Introduction}
\label{sec:introduction}

Observational studies at various wavelength regimes have converged
towards a widely accepted galaxy evolution picture. In this scenario
galaxies are thought to evolve in time from an initial star-formation
dominated state with blue optical colors into the most massive
`red-and-dead' galaxies \citep{bell04, bell04dusty, borch06, faber07,
  brown07, hopkins07}.  The transitional phase, that links the blue
and red galaxy states, is reflected in a sparsely populated region in
color-magnitude diagrams that is referred to as the `green valley'.
On the other hand, past cosmological models have been unsuccessful in
describing the formation of massive galaxies
\citep[e.g.][]{white78,white91}. They have led to systematic
over-predictions of the number of both massive blue, and the most
massive red galaxies in the universe. Only recently have these
problems been overcome by implementing `AGN feedback' in the models
\citep{croton06, bower06, hopkins06, sijacki06, sijacki07}. Two types
of AGN feedback are often invoked. The first is known as the `quasar'
or `truncation' mode. In this mode quasar winds are thought to quench
star formation, and cause galaxies to fade to red colors by expelling
a fraction of the gas from the galaxy.  The second, often referred to
as `radio' or 'maintenance' mode is linked to radio AGN outflows in
already massive, red galaxies. These outflows are thought to heat the
surrounding medium, and thereby prevent further star formation in the
galaxy and hence limit growth from creating overly high mass
galaxies. Although observational evidence supporting AGN feedback is
growing \citep{best06,bundy08,nesvadba08,smo09}, its impact on galaxy
formation and evolution is still poorly understood. Here we focus on
studying the link between radio AGN activity and massive galaxy
formation.

Studies of radio AGN suggest that powerful (\lum~$\gtrsim10^{25}$~\wh
) and weak (\lum~$<10^{25}$~\wh ) radio AGN represent different galaxy
populations \citep[e.g.][]{fr74, ledlow96, kauffmann08}. Based on a
sample of local radio AGN, \citet{kauffmann08} have shown that those
with strong emission lines of high-ionization potential species are
predominantly powerful in radio.
As has been suggested in a model developed by \citet{hardcastle06}
this high/low excitation classification may represent a principal separator between
populations fundamentally different in their black hole accretion
mechanisms (see also \citealt{evans06,allen06,kewley06}). In this model,
central super-massive black holes of high-excitation radio AGN (HERAGN
hereafter) accrete in a standard (radiatively efficient) way from the
cold phase of the intra-galactic medium (IGM), while those of
low-excitation radio AGN (LERAGN hereafter) are powered in a
radiatively inefficient manner by Bondi accretion of the hot
IGM. Although still awaiting a robust confirmation, this model
successfully explains many observed properties of radio AGN.

Recently, \citet{smo09} have used a unique sample of weak (VLA-COSMOS;
\citealt{schinnerer07, smo08}) radio AGN, that reaches out to $z=1.3$,
to study their host galaxies, and cosmic evolution.  They find that,
i) already at $z\sim1$, weak radio AGN occur in red-sequence galaxies
with the highest stellar and black hole masses, and ii) contrary to
powerful radio AGN, the volume-averaged number-density of weak radio
AGN evolves only modestly with cosmic time (see \citealt{smo09} and
references therein). Based on these results, they have proposed an
evolutionary scenario in which powerful (predominantly
high-excitation) and weak (predominantly low-excitation) radio AGN
represent different (i.e.\ earlier and later, resp.) stages of the
blue-to-red galaxy evolution. This scenario suggests that the
triggering of radio AGN activity is a strong function of host galaxy
properties, linked to different stages of massive galaxy
formation. It  also illuminates the mechanisms of `AGN
feedback', regularly invoked in cosmological models, as it suggests
that powerful and weak radio AGN activities are bound to different types
of feedback.

The main goal of this Letter is to test the above outlined models and
scenarios. To do this we utilize a large local sample of radio AGN
(\s{sec:data} ), in which accurate low/high excitation classifications
can be obtained. In \s{sec:hostprops} \ we explore the fundamental
differences between these types of radio AGN, and we put them into a
perspective of massive galaxy build-up in \s{sec:discussion} .

\section{The FIRST-NVSS-SDSS radio AGN sample}
\label{sec:data}

In order to investigate the principal differences between various
radio AGN types, we make use of a unified catalog of radio objects
detected by NVSS, FIRST, WENSS, GB6, and SDSS \citep{kim08}.  We have
augmented this catalog with derivations of emission line fluxes,
4000\AA\ breaks, stellar masses, and stellar velocity dispersions,
drawn from the SDSS-DR4 `main' spectroscopic sample (see
\citealt{kauffmann03a,kauffmann03b} and reference
therein)\footnote{The catalogs with emission line fluxes
  (emission$\_$lines$\_$v5.0$\_$4.fit), and stellar mass, Dn(4000),
  and stellar velocity dispersions (agn.dat$\_$dr4$\_$release.v2) have
  been downloaded from {\tt www.mpa-garching.mpg.de/SDSS/DR4/}}. The
cross-correlation yielded 20,648 radio object entries in the catalog
with available line flux measurements\footnote{This catalog is
  available at {\tt
    www.astro.caltech.edu/$\sim$vs/RadioCat.php}}. Following
\citet{kim08} we further limit this catalog to 6,640 unique objects
that have been detected by both the FIRST and NVSS surveys at 20~cm
(equivalent to sample `C' in their Tab.~8, but without the
`overlap~$=$~1' criterion).

In order to access the most accurate emission line flux estimates we
restrict the above defined sample of unique objects in redshift to
$0.04<z<0.1$, and select only sources with stellar mass, and
4000\AA\ break measures\footnote{This includes all galaxies with
  available stellar velocity dispersion, and O[III] luminosity
  estimates.}.  We define emission-line galaxies as those where the
relevant emission lines (H$\alpha$, H$\beta$, O[III,$\lambda$5007],
N[II,$\lambda$6584]) have been detected at $\mathrm{S/N}\geq3$, and
consider all galaxies with $\mathrm{S/N}<3$ in these lines as
absorption line systems.  Given that the latter are luminous at 20~cm,
they can be considered to be (low excitation) AGN (see
e.g.\ \citealt{best05,smo08} for a more detailed discussion).
Further, using standard optical spectroscopic diagnostics
\citep{bpt81,kauffmann03b, kewley01, kewley06} we sort the
emission-line galaxies into 1) star forming, 2) composite, 3) Seyfert,
and 4) LINER galaxies (see \f{fig:bpts} ).  The last two classes have
been selected by requiring $\mathrm{S/N}\geq3$ in
S[II,$\lambda$$\lambda$6717,6731] and [OI, $\lambda6300$], as well as
`unambiguously' by imposing combined criteria using 3 emission line
flux ratios (see middle and right panels in \f{fig:bpts} ).

\newpage
\begin{figure}
\center{
\includegraphics[bb = 30 450 500 640, width=\columnwidth]{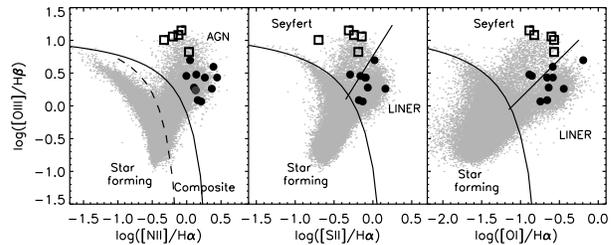}
\caption{ Optical spectroscopic diagnostic diagrams (see
  \citealt{kauffmann03b, kewley06}) that separate emission-line
  galaxies into star forming,
  composite galaxies, and various types of AGN (Seyferts and
  LINERs). Small gray dots represent galaxies from the SDSS DR4
  ``main'' spectroscopic sample. Large open squares (filled dots)
  denote $z<0.1$ 3CRR radio galaxies independently classified (based
  on their core X-ray emission) as systems with radiatively efficient
  (inefficient) black hole accretion \citep{evans06}.
  \label{fig:bpts}}
}
\end{figure}

The spectroscopic selection yields final samples of $\sim310$ star
forming galaxies and $\sim480$ AGN, out of which $\sim110$ are
`unambiguous' Seyferts, $\sim120$ are `unambiguous' LINERs, and
$\sim250$ are absorption line systems. Hereafter, we will separately
analyze the properties of absorption line AGN, LINERs, and Seyferts,
keeping in mind that the first two are low-excitation systems, while
the last has luminous and hard UV sources which ionize the observed
emission lines \citep[e.g.][]{kewley06}.  The full range of 20~cm
radio power\footnote{Computed using the NVSS total flux densities
  ($F_\nu$), and assuming a spectral index of $\alpha=0.7$
  ($F_\nu\propto\nu^{-\alpha}$)} in these samples is $\sim
10^{22}-10^{25}$~\wh .

\section{Radio AGN host galaxy properties}
\label{sec:hostprops}

In \f{fig:props} , we show the distributions of color, 4000\AA\ break
strength [$\mathrm{D_n(4000)}$] and stellar mass for the host galaxies
of radio luminous AGN, spectroscopically divided into Seyfert, LINER,
and absorption line galaxies. While LINER and absorption line systems
follow very similar distributions (as expected if low-excitation
systems form one family), there are obvious differences compared to
Seyfert galaxies. The former on average have redder optical $g-r$
colors (left panel in \f{fig:props} ), larger values of
$\mathrm{D_n(4000)}$ (implying older stellar populations; middle
panel), and higher stellar masses (left panel).  From these plots it
is obvious that Seyfert galaxies represent a population with
properties in an intermediate range, linking those of star forming
galaxies at one extreme to those of massive red galaxies at the
other. In particular, Seyfert galaxies in the FIRST-NVSS-SDSS sample
are consistent with `green valley' objects, a population thought to be
in transition from star formation dominated- to a `red-and-dead'-
state \citep[e.g.][]{faber07}. We will discuss this further in
\s{sec:discussion} .

In order to investigate the central super-massive black hole
properties of radio AGN, in \f{fig:bhprops} \ we show the distribution
of stellar velocity dispersion ($\sigma$), and the ratio of O[III]
luminosity and $\sigma^4$ for the LINERs and Seyferts in the
FIRST-NVSS-SDSS sample. As black hole mass is tightly correlated with
the bulge stellar velocity dispersion \citep[$\log{M_\mathrm{BH}} =
  8.13 +
  4.02\log{\frac{\sigma}{200~\mathrm{km\,s^{-1}}}}$;][]{tremaine02}
$\log{\sigma}$ is a proxy for black hole mass. Furthermore, as
$L_\mathrm{O[III]}$ traces well the AGN bolometric luminosity (see
\citealt{heckman04} for details), the quantity
$L_\mathrm{O[III]}/\sigma^4$ is proportional to the mass accretion
rate onto the black hole (in Eddington units;
$L/L_\mathrm{EDD}=\dot{M}_\mathrm{BH}/\dot{M}_\mathrm{EDD}$).
\f{fig:bhprops} \ shows that radio luminous LINERs have systematically
higher black hole masses, and lower accretion rates (in Eddington
units) than radio luminous Seyferts. This is consistent with the
overall properties of LINER and Seyfert galaxies \citep{ho05,
  kewley06}, and in support of the radio AGN scenario developed by
\citet{hardcastle06}, and based on the results presented in
\citet[][E06 hereafter]{evans06}.

Using high resolution X-ray imaging of radio cores in a sample of 22
3CRR ($z<0.1$) radio galaxies, E06 have studied accretion rates in
various types of radio AGN. They have found that X-ray cores of
usually powerful, edge brightened (FR~II) radio galaxies are dominated
by absorbed, accretion related emission. Contrary to this, they find
that X-ray cores of typically weak, core dominated (FR~I) radio
galaxies likely arise from a jet, and any additional accretion-related
components in these have low radiative efficiencies.  Based on these
results, \citet{hardcastle06} have assumed that it is the excitation
state (rather than radio morphology) that is the principal separator
of the black-hole accretion mechanism in radio AGN. Recently,
high-quality optical spectroscopy, that allows a robust high/low
excitation classification of the E06 sample, has become
available. \citet{buttiglione09} have computed emission line fluxes
for a large number of radio AGN, including most of the galaxies in the
E06 sample. Thus, it is now possible to view the E06 sample robustly
separated into low-excitation (LINER) and high-excitation (Seyfert)
state systems. In \f{fig:bpts} \ we have overplotted the 3CRR galaxies
studied in E06 with available spectroscopic line measurements. The
properties of this sample are summarized in \t{tab:3CRR}
. \f{fig:bpts} \ clearly shows that the 3CRR radio galaxies
independently classified as radiatively inefficient accretors (E06)
are LINERs, while those identified as accreting efficiently are
Seyferts (see also last column in \t{tab:3CRR} ). Only two galaxies
(3C~31, and 3C~449) are classified here as Seyferts, yet they have been
identified as radiativelly inefficient accretors by E06. However, as
evident from \t{tab:3CRR} \ their black hole accretion efficiencies
may also be consistent with radiativelly efficient accretion.
Overall, the black hole masses of the 3CRR LINER galaxies
($<\log{M_\mathrm{BH}}> = 9.14\pm0.06 $~\msol ) are systematically
higher than those of 3CRR Seyferts ($<\log{M_\mathrm{BH}}> = 8.3 \pm
0.1$~\msol ), consistent with the black hole properties of
FIRST-NVSS-SDSS LINERs and Seyferts.

\begin{deluxetable*}{c|c|c|c|c|c|c|c}
\tablecaption{Properties of 3CRR radio AGN
\label{tab:3CRR}
  }
\tablehead{
\colhead{name} &
\colhead{redshift} &
\colhead{N[II]/H$\alpha$} &
\colhead{S[II]/H$\alpha$} &
\colhead{O[I]/H$\alpha$} &
\colhead{O[III]/H$\beta$} &
\colhead{$\log{M_\mathrm{BH}}$} &
\colhead{$L_\mathrm{0.5-10keV}/L_\mathrm{EDD}$}
}
\startdata
3C 33 &  0.060 &  0.630 &  0.720 &  0.250 & 11.452 &  8.68 & $1.6\times10^{-03}$   \\
3C 98 &  0.030 &  0.760 &  0.570 &  0.150 & 12.040 &  8.23 & $3.7\times10^{-04}$   \\
3C 390.3 &  0.056 &  0.470 &  0.200 &  0.270 & 10.125 &  8.53 & $1.1\times10^{-02}$   \\
3C 403 &  0.059 &  0.840 &  0.490 &  0.130 & 14.160 &  8.41 & $3.3\times10^{-03}$   \\
3C 452 &  0.081 &  1.080 &  0.650 &  0.270 &  6.652 &  8.54 & $3.3\times10^{-03}$   \\ 
\hline\hline
3C 31 &  0.017 &  0.990 &  0.690 &  0.140 &  2.867 &  7.89 & $<2.0\times10^{-04}$   \\
3C 449 &  0.017 &  1.380 &  0.510 &  0.130 &  3.000 &  7.71 & $<7.0\times10^{-03}$   \\
\hline
3C 66B &  0.022 &  2.450 & -- &  0.260 &  3.955 &  8.84 & $<4.4\times10^{-05}$   \\
3C 83.1B &  0.027 &  1.350 & -- & -- &  1.736 &  9.01 & $<6.8\times10^{-06}$   \\
3C 84 &  0.018 &  1.120 &  1.050 &  0.640 &  4.976 &  9.28 & $<9.2\times10^{-06}$   \\
3C 264 &  0.022 &  1.450 &  0.660 &  0.220 &  1.222 &  8.85 & $<1.8\times10^{-05}$   \\
3C 272.1 &  0.004 &  1.280 &  0.860 &  0.230 &  1.900 &  9.18 & $<8.5\times10^{-07}$   \\
3C 274 &  0.004 &  2.320 &  1.450 &  0.360 &  1.824 &  9.38 & $<4.3\times10^{-07}$   \\
3C 296 &  0.025 &  1.840 &  0.810 &  0.220 &  2.700 &  9.13 & $<1.2\times10^{-05}$   \\
3C 338 &  0.032 &  1.630 &  0.740 &  0.180 &  1.167 &  9.23 & $<2.0\times10^{-05}$   \\
3C 465 &  0.030 &  2.770 &  0.790 &  0.260 &  2.706 &  9.32 & $<2.2\times10^{-04}$ 
\enddata
\tablenotetext{.}{The first column represents the 3CRR source. The
second column shows the redshift, and columns (3) to (6) line flux
ratios for each source (shown in \f{fig:bpts} \ and adopted from
\citealt{buttiglione09}). The last two columns, adopted from
\citet[E06]{evans06}, represent the black hole mass, and accretion
efficiency (in Eddington units) for each AGN (see text for details;
the upper limits are obtained assuming $N_H=10^{24}$~atoms~cm$^2$; see
E06). Radiatively efficient (top) and inefficient (bottom) accretors,
as defined by E06 based on their core X-ray emission, are separated by
the double horizontal line. The single horizontal line separates
Seyferts (top) and LINERs (bottom), identified here based on their
optical spectroscopic properties (see \f{fig:bpts} , and text for
details).}
\end{deluxetable*}

\begin{figure}
\center{
\includegraphics[bb = 54 590 486 742, width=\columnwidth]{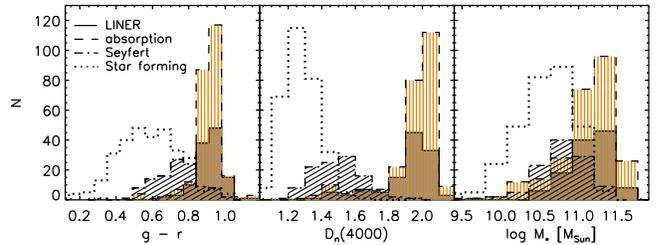}
\caption{ Distribution of $g-r$ (left), 4000\AA\ break (middle), and
  logarithm of stellar mass (right) for radio luminous LINERs (filled
  histograms), absorption line systems (vertically hatched
  histograms), and Seyferts (diagonally hatched histograms), drawn
  from the FIRST-NVSS-SDSS sample. For comparison, the distribution of
  radio luminous star forming galaxies is also shown. Note that
  Seyfert, i.e.\ high-excitation, radio sources occupy an intermediate
  region between star forming and low-excitation (LINERs, and
  absorption line) AGN in these diagrams.
  \label{fig:props}}
}
\end{figure}

\begin{figure}
\center{
\includegraphics[bb = 54 400 486 572, width=\columnwidth]{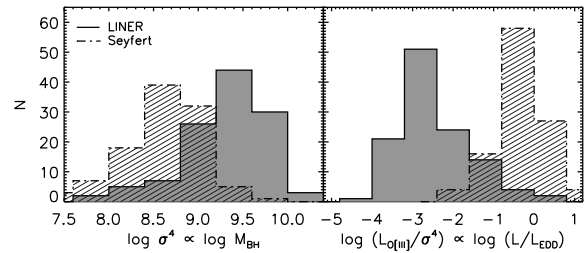}
\caption{ Distribution of a) velocity dispersion, $\sigma^4$ 
  (left panel), which is proportional to black hole mass
  \citep{tremaine02}, and b) $\log{L_\mathrm{O[III]}/\sigma^4}$ (right
  panel), proportional to black hole accretion rate \citep[in
    Eddington units;][]{heckman04} for radio luminous LINERs (filled
  histograms), and Seyferts (hatched histograms) studied here.
  \label{fig:bhprops}}
}
\end{figure}

\section{Summary and discussion}
\label{sec:discussion}

We have based this study on a local ($0.04<z<0.1$) 20~cm selected
sample of radio AGN detected by the FIRST and NVSS surveys. By
cross-matching this sample with optical spectroscopic measurements
from the SDSS survey, we have investigated the principal host galaxy
and central super-massive black hole differences between various types
of radio AGN.  In the following we will refer to the radio LINER and
absorption line AGN as low-excitation radio AGN (LERAGN), and to
Seyferts as high-excitation radio AGN (HERAGN). Furthermore, we take
the powerful (\lum~$\gtrsim10^{25}$~\wh ) and weak
(\lum~$<10^{25}$~\wh ) radio AGN to be predominantly high- and
low-excitation systems, respectively (based on the results from
\s{sec:introduction} ).

\subsection{The black hole properties of radio AGN: A
  low/high excitation state dichotomy} 

Based on the FIRST-NVSS-SDSS sample we have shown that the mass
accretion rates (in Eddington units) in low-excitation radio AGN are
on average substantially lower than in high-excitation radio AGN,
while the trend is opposite for their black hole masses. This is
consistent with the high-resolution X-ray analysis of the cores of
local 3CRR radio galaxies (E06). Combining these with recent line flux
measurements \citep{buttiglione09} suggests that LERAGN accrete onto
their black holes in a radiatively inefficient way, while the black
hole accretion in HERAGN occurs via a standard thick disk in a
radiatively efficient manner. Thus, the division of radio AGN into
low- and high-excitation sources seems to be a good proxy for
identifying radio AGN, fundamentally different in their central
supermassive black hole properties. Furthermore, we have shown that
these two types of radio AGN show systematic differences also in their
host galaxy properties on large (kpc) scales.  As outlined below, this
link between pc-scale and kpc-scale galaxy properties may be explained
if HE- and LE-RAGN represent different stages in the process of
massive red galaxy formation.

\begin{figure}[t!]
\center{ \includegraphics[bb = 79 400 410 757,
    width=\columnwidth]{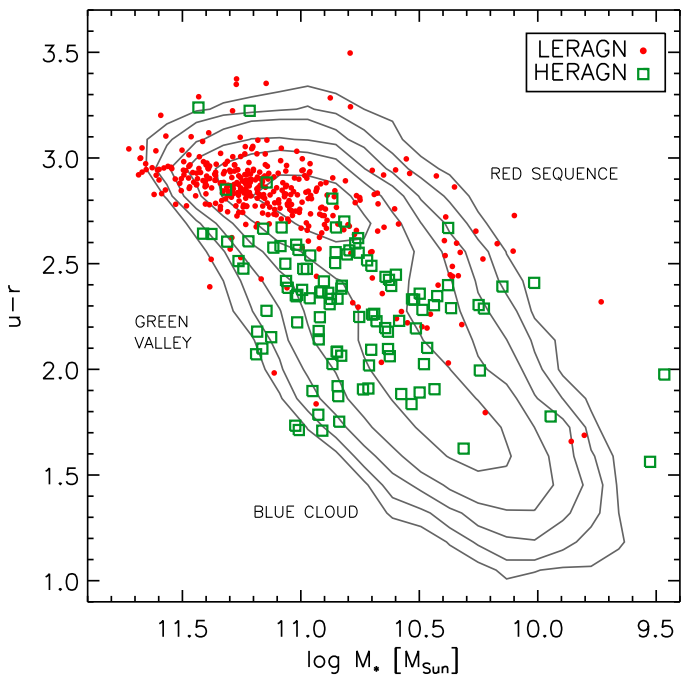}
\caption{ $u-r$ color versus stellar mass for the general population
  of galaxies ($0.04<z<0.1$; drawn from the sample used in
  \citealt{smo06}) shown in contours. The contour levels start at 26,
  and continue in steps of $2^n$ ($n=1,2...5$). The distributions of
  low- and high- excitation radio AGN (LERAGN and HERAGN respectively)
  is also shown (filled and open symbols, resp.). Note that HERAGN
  represent a transition population, consistent with `green valley'
  objects, while LERAGN predominantly occupy the red sequence.
  \label{fig:cmd}}
}
\end{figure}

\subsection{The low/high excitation radio AGN dichotomy in the context
  of massive galaxy formation}  

Galaxies are thought to evolve in time from an initial stage with
irregular or spiral morphology and blue optical colors towards
elliptical morphologies with red optical colors and the highest
stellar masses ($M_*\gtrsim10^{11}$~\msol ;
e.g.\ \citealt{faber07}). In the context of this red massive galaxy
formation picture, the results presented here imply that LERAGN and
HERAGN represent different stages of this process. As summarized in
\f{fig:cmd} , low-excitation radio AGN are predominantly hosted by the
most massive galaxies, with the reddest colors, consistent with final
stages of massive red galaxy formation. On the other hand, the host
galaxies of high-excitation radio AGN have colors consistent with a
transitional region (`green valley') between blue and red galaxies
(see also \s{sec:hostprops} ). Compared to LERAGN, they on average
have younger stellar populations, lower black-hole masses, and higher
accretion rates (see \s{sec:hostprops} ). Such properties are
consistent with an intermediate, very active phase in which galaxies
are undergoing substantial stellar and black hole mass build-up on
their evolutionary path towards a massive `red-and-dead' state.

Our results are in agreement with the scenario presented by
\citet{smo09}. Based on a study of the evolution of weak (VLA-COSMOS)
and powerful (3CRR, 6CE, and 7CRS; \citealt{willott01}) radio AGN
since $z=1.3$, they have proposed that powerful (high-excitation)
radio AGN represent an intermediate stage in the formation of a
massive-red galaxy, while weak (low-excitation) radio AGN occur in the
latest phases of this process when the galaxy has already assembled
both its stellar and black hole masses. Thus, in the context of AGN
feedback as a relevant mechanism for the formation of massive
galaxies, only LERAGN, i.e.\ weak (\lum~$<10^{25}$~\wh ) radio AGN,
are expected to contribute to the so called `radio mode' feedback,
responsible for limiting stellar mass growth. Namely, in cosmological
models this mode becomes important when a galaxy has already assembled
most of its mass, and it has formed a hot spherical halo. Such
properties are observed in massive galaxies that host predominantly
{\em weak} radio AGN.  On the other hand, the results presented here
show that HERAGN, i.e. powerful (\lum~$>10^{25}$~\wh ) radio AGN,
occur in a `transitional' evolutionary state linking blue star forming
and massive red galaxies. Thus, in terms of AGN feedback, powerful
radio AGN outflows are more closely linked to a different feedback
phase, namely the `quasar mode' thought to be relevant for star
formation quenching in galaxies that causes blue star-forming galaxies
to fade to redder colors.

\acknowledgments VS thanks Amy Kimball for help with the Unified Radio
Catalog, as well as Nick Scoville and Scott Schnee for insightful
comments on this manuscript, and Dominik A.\ Riechers for help with
choosing `proper' colors in the last figure. The National Radio
Astronomy Observatory is a facility of the National Science Foundation
operated under cooperative agreement by Associated Universities, Inc.

{}

\end{document}